\documentclass[authorversion,sigconf,nonacm]{acmart} 

\usepackage{multirow} 
\usepackage{graphicx}
\usepackage{array}
\usepackage{boldline}
\usepackage{soul}
\usepackage{microtype}
\AtBeginDocument{%
  \providecommand\BibTeX{{%
    \normalfont B\kern-0.5em{\scshape i\kern-0.25em b}\kern-0.8em\TeX}}}

\setcopyright{acmlicensed}
\copyrightyear{2024}
\acmYear{2024}
\setcopyright{acmlicensed}\acmConference[CHI '24]{Proceedings of the CHI Conference on Human Factors in Computing Systems}{May 11--16, 2024}{Honolulu, HI, USA}
\acmConference[CHI '24]{Proceedings of the CHI Conference on Human Factors in Computing Systems}{May 11--16, 2024}{Honolulu, HI, USA}
\acmBooktitle{Proceedings of the CHI Conference on Human Factors in Computing Systems (CHI '24), May 11--16, 2024, Honolulu, HI, USA}
\acmDOI{XXXXXXX.XXXXXXX}
\acmISBN{XXX-X-XXXX-XXXX-X/XX/XX}

\begin{document}

\title[Probing Design Opportunities for SEL Agents through Children's Peer Co-Creation of Social-Emotional Stories]{``My lollipop dropped…''—Probing Design Opportunities for SEL Agents through Children's Peer Co-Creation of Social-Emotional Stories}
\titlenote{To appear at ACM CHI EA '24.} 

\author{Hanqing Zhou}
\orcid{0009-0004-2077-6030}
\affiliation{%
  \institution{School of Design, SUSTech}
  \city{Shenzhen}
  \country{China}
}
\email{12331483@mail.sustech.edu.cn}

\author{Anastasia Nikolova}
\orcid{0009-0001-7476-1596}
\affiliation{%
  \institution{Data Science Department, Duke Kunshan University}
  \city{Kunshan}
  \country{China}
}
\email{anastasia.nikolova@duke.edu}

\author{Pengcheng An}
\authornote{Corresponding Author.}
\orcid{0000-0002-7705-2031}
\affiliation{%
  \institution{School of Design, SUSTech}
  \city{Shenzhen}
  \country{China}
}
\email{anpc@sustech.edu.cn}

\renewcommand{\shortauthors}{H. Zhou, A. Nikolova \& P. An}

\begin{abstract}
This Late-Breaking Work explores the significance of socio-emotional learning (SEL) and the challenges inherent in designing child-appropriate technologies, namely storytelling agents, to support SEL. We aim to probe their needs and preferences regarding agents for SEL by conducting co-design which involves children co-creating characters and social emotional stories. We conducted collaborative story-making activities with children aged four to six years old. Our findings could inform the design of both verbal and nonverbal interactions of agents, which are to be aligned with children’s understanding and interest. Based on the child-led peer co-design, our work enhances the understanding of SEL agent designs and behaviors tailored to children’s socio-emotional needs, thereby offering practical implications for the more effective SEL tools in future HCI research and practice.
\end{abstract}

\ccsdesc[500]{Human-centered computing~Empirical studies in HCI}

\keywords{Social emotional learning, Co-design, Collaborative storytelling, Agent, Generative AI}

\begin{teaserfigure}
  \centering
  \includegraphics[width=\textwidth]{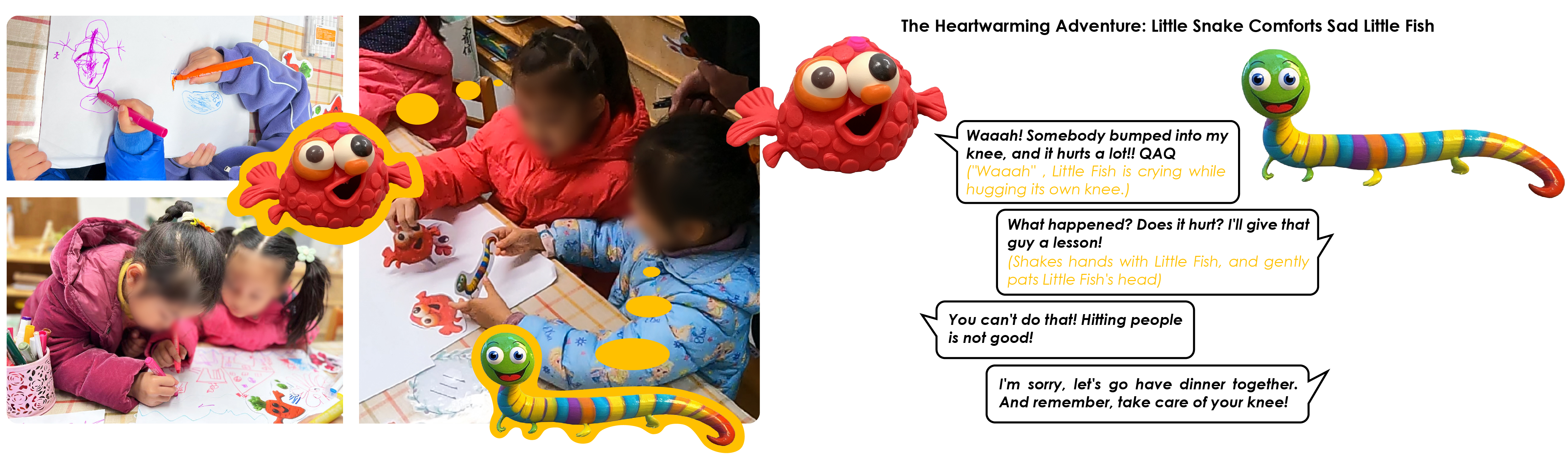}
  \caption{Collaborative storytelling and exemplar socio-emotional narratives by children.}
  \label{fig:teaser}
\end{teaserfigure}

\maketitle

\newpage

\section{Background}
Social-emotional learning (SEL) is considered an important foundation for children's emotional intelligence. SEL encompasses a variety of skills, including emotion processing, empathy demonstration, positive relationship establishment, and responsible decision-making \cite{jones2017promoting}. Research consistently demonstrates that well-developed socio-emotional skills in children are associated with higher academic performance, healthier social interactions, and lower levels of mental illness \cite{durlak2011impact,murano2020meta}. Furthermore, SEL provides children with tools to navigate social interactions and personal feelings to adapt to everyday challenges \cite{durlak2011impact}. This early investment in emotional and social learning lays the foundation for empathetic, competent, and flexible adults \cite{durlak2011impact}. Realizing the significance of SEL, it becomes essential to explore novel ways and educational methods that can develop these skills in children.

Despite its importance, designing child-appropriate technologies to support SEL is a complex challenge. Prior Human-Computer Interaction (HCI) studies highlight the potential of grounding the "participatory turn" in AI design \cite{delgado2023participatory} and conversational agents in enhancing children's SEL \cite{sanoubari2021robots}, some studies have revealed that conversational agents can offer a new mechanism for facilitating parent-child bonding by storytelling using Al to adapt to dynamic user needs \cite{zhang2022storybuddy}, and some of them have found their effectiveness largely depends on aligning the agents' interactions with children's preferences and understanding \cite{sanoubari2021robots}. The challenge lies in accurately identifying these preferences, including both the verbal and nonverbal aspects of agents \cite{sanoubari2021robots}. Co-design methods that actively involve children have been shown to effectively reveal their unspoken needs and latent preferences toward emerging technologies \cite{warren2022lessons}. We thus believe leveraging co-design with children can generate instrumental design inputs for crafting storytelling agents' interactions aligned with children's understanding and interest, and are suitable for their SEL.

In this late-breaking pilot study, we set out to gather inputs for creating child-appropriate storytelling agents to support their SEL by organizing children’s co-design activities in which they created and role-played SEL stories. Studies in other forms of technology design, such as educational apps and tangible interfaces, show that co-design activities including collaborative storytelling and role-playing, could not only boost children's creativity and self-expression but also create opportunities to embody their SEL \cite{sanusi2023preparing}. When children are active participants in creating stories and role-playing scenarios, they engage more deeply with the concepts of empathy, emotional recognition, and expression–key components that could shed light on their SEL process \cite{paracha2019co, liu2024he}. Role-playing activities also have been demonstrated to be effective in facilitating children to explore and express emotions \cite{paracha2019co, liu2024he}. Additionally, the integration of AIGC tools into the process of creative expression development and co-ideation process, especially for children, has been a subject of exploration of the research \cite{han2023design, liu2024he}. 
The studies have highlighted the potential of such tools facilitating children’s active participation in design activities and providing immediate visualization, thereby enhancing their creativity skills and expression of needs and preferences in the process \cite{han2023design}. Nonetheless, it is crucial to ensure that these activities are thoughtfully designed in order to more pertinently gather insights into children's SEL \cite{sanusi2023preparing}. By integrating these participatory design approaches, we aim to extract rich and authentic implications about how we could design child-appropriate SEL storytelling agents whose verbal and nonverbal interactions align with children's understanding and interests and are suitable for their SEL.

Namely, our work adopted a co-design approach targeted at embodying children's SEL through peer-to-peer storytelling \cite{sanoubari2021robots}. Moreover, previous research has demonstrated the development of correspondence between non-verbal and verbal behavior in preschool children \cite{risley1968developing,lovaas1961interaction,childers2003children,pelletier2004action}, so in our process, we focused on both the verbal and nonverbal elements within their natural interactions and emotional expressions \cite{warren2022lessons, pelletier2004action} enacted in their co-storytelling, to supplement the previous emphasis on the verbal aspect of agent design \cite{alves2017yolo}. We employed a blend of AIGC tools and tangible materials to create agent character images that resonate with their preferences and engaged children in collaborative storytelling through role-playing and art-making using materialized AI image cutouts. Children’s emotionally rich narratives about the agent's experiences were utilized to identify the preferred characteristics and features of the agent. Additionally, we collected the stories’ contexts and agent’s responses to the emotional events—both verbal and nonverbal—created by children in collaborative story-making sessions.
Our preliminary study aims to inform future HCI exploration by providing insights for designing child-appropriate storytelling agents. These insights have been gathered from the child-led peer-to-peer storytelling activities we presented. With a focus on translating the co-created story characters and social-emotional conversations into design inputs for future verbal and nonverbal interactions with SEL agents, we contribute a set of design implications with concrete examples from children's co-design outcomes, emphasizing the importance of fostering emotional inquiry, enabling collaborative solutions for negative emotions, incorporating varied interactions (both verbal and nonverbal elements) for emotional connection and so on.

\begin{figure*}[tb]
  \centering
  \includegraphics[width=\textwidth]{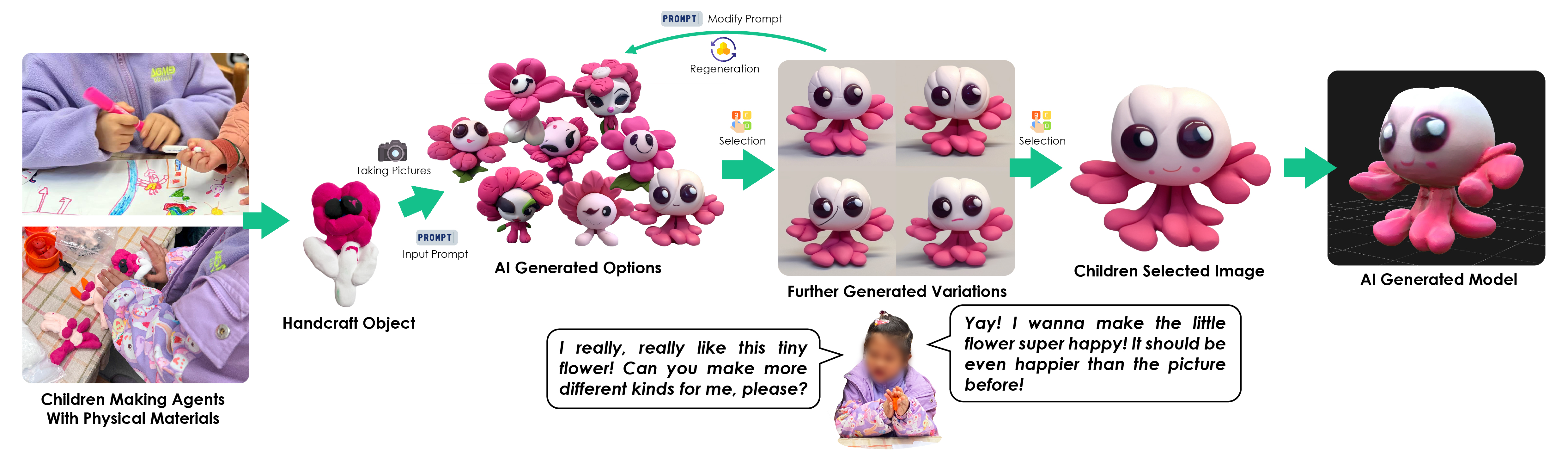}
  \caption{Illustration for child-AI co-creating: integrating physical materials and generative AI in the creative process.}
  \label{Character}
\end{figure*}

\section{Methodology}

\subsection{Participants}

In our study, we recruited 8 children with an average age of 5.5 years (four six-year old, three five-year old and one four-year old), comprising 6 girls and 2 boys from various classes in the same kindergarten. 
Participants were randomly recruited with the help of the teachers. 
Our protocol obtained approval from the institutional research ethics board, and for confidentiality, all names in this paper are substituted with pseudonyms. 
Parents were provided with an informed consent form outlining the activity, and their review and signature were required before their child's participation.

\subsection{Design process}
We conducted two co-design sessions, each lasting approximately 1.5 hours. 
Recognizing the potential of various materials to stimulate children's creativity \cite{sari2021can,eriksson2021glitter}, we prepared several types of materials for hands-on creations before each session. This included physical items like pencils, crayons, and clay, as well as the incorporation of the widely-used image-generative AI tool, Midjourney \cite{midwebsite}. This decision was based on its advantages, such as lowering the threshold for art-making and visual expressions, offering customization and variation to empower children's creativity \cite{radhakrishnan2023midjourney}, and rapid, efficient image generation based on simple text and image prompts \cite{radhakrishnan2023midjourney}.

Children's handcrafted artifacts and their verbal description about their creation were used as image and text prompts, to generate AI character images during activities. We prepared template prompts following the guidelines on Midjourney prompt writing \cite{midGuidelines}, tailored to our specific case. We utilized a format of "Image URL + Subject + Subject Descriptor + Style Descriptor + White Background + Parameters (Aspect Ratio: 16:9)." The Image URL is the link to the image input created by children. The prompt's subject, subject descriptor, and style descriptor elements are derived from the verbal expressions of children while explaining their artifacts. Detailed examples of these guidelines can be found in the APPENDIX.

\subsubsection{Co-design Session 1: Agent Creating and Sharing} The first co-design study included 3 key phases: Warm-up Activities, Agent Creating Activities, and Sharing Activities.

\textbf{Warm-up Activities:} The session began by encouraging children to familiarize themselves with the materials on the table and share their crafting experiences. The activity took place in the kindergarten's handicraft classroom, children were allowed to explore and choose their preferred crafting tools and materials, with the teacher's facilitation.

As tangible inspirations, two dolls were introduced by the researcher—one with a simple responsive function (repeating what the child said and shaking its ears) and the other without any interactive features. Children took turns playing with both dolls, and we asked about the presence of similar companion dolls in their lives, exploring when these dolls are engaged with and how they were experienced. The goal was to understand if children would be willing to share their daily life and emotional experiences with such companions, uncovering potential opportunities for designing agents that support children's social-emotional learning.

\textbf{Agent Creating Activities:} In order to gather design inspirations for child-preferred agents, we actively accumulated ideas and examples directly from the children. During the activities, children used AI-infused art-making materials to design and craft characters (a detailed process is illustrated in \autoref{Character}).
Children took the lead in the entire creative process, primarily starting from clay kneading or painting to create initial artwork, shown on the left side of \autoref{Character}.
A researcher helped children take photos of their physical artwork as the image input for generative AI. 

Meanwhile, children were invited to verbally explain their artifacts and their articulation was utilized by the researcher as the text prompt for generative AI on the spot.
They could modify their artifact or iterate their verbal articulation as updated AI inputs for regeneration, aiming for the most suitable visual expression in both content and art style. 
They were also encouraged to give further comments for the researcher to generate more variations with AI tool or make minor adjustments using retouching tools, as depicted in the middle of \autoref{Character}.
This combination of digital and physical materials aimed to lower the creative threshold for children, enabling effective shaping of ideas and enhancing engagement. 
The finalized agent design could be transformed from a 2D mage into a 3D shape by another AI tool (see the right side of \autoref{Character}).
 


\textbf{Agent Design Sharing Activities:} Following the creation session, the eight children each took turns to present the characters they crafted to the others. This presentation included sharing visual and functional attributes, along with explaining the desirable characteristics and interactive features of the agents they envisioned. Based on a group discussion, the participants collectively identified four characters that received unanimous approval, agreeing to use them in the next co-design session for co-creating social-emotional stories.

\subsubsection{Co-design Session 2: Collaborative Story-making and Sharing}
\begin{figure*}[tb]
  \centering
  \includegraphics[width=\textwidth]{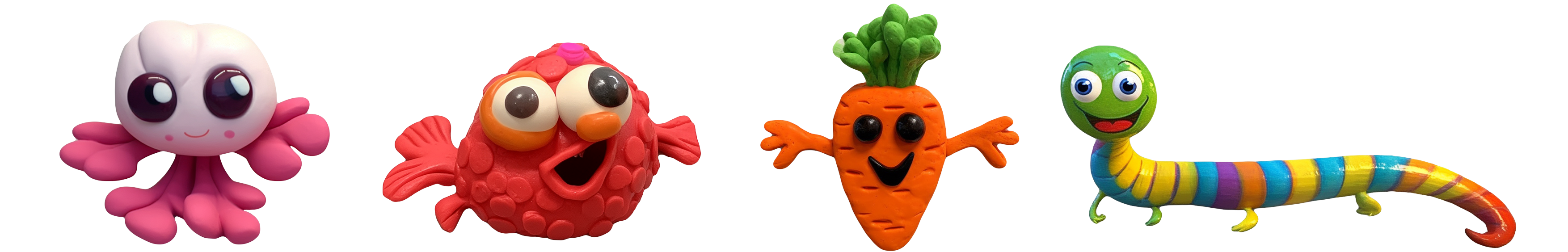}
  \caption{The four characters selected by the children in Co-design Session 1}
  \label{SELECTED CHARACTER}
\end{figure*}

This session centered on developing and enacting stories and scenarios of social-emotional events. This approach, adopted from \cite{liu2024he}, showed itself an effective method to elicit consistent and emotionally resonant responses that can be used for agents, aligning with the children's experiences and perspectives. The four characters selected by the children in Co-design Session 1 (as depicted in \autoref{SELECTED CHARACTER}) were printed and cut out before the activities, offering tangible materials for collaborative storytelling.

\begin{table*}[ht]
\caption{Overview of Selected Emotions and Characters during Collaborative Story-making.}
\label{tab:choice}
\centering
\begin{tabular}{l|ll|ll|ll|ll}
\toprule
Selected Emotions & \multicolumn{2}{c|}{Angry} & \multicolumn{2}{c|}{Sad} & \multicolumn{2}{c|}{Happy} & \multicolumn{2}{c}{Scared} \\ \hline
Name & Tom & Bob & Kate & Betty & July & Mia & Jane & Xixi\\ \hline
Characters used in story-making & \multirow{2}{*}{\begin{tabular}[c]{@{}l@{}}{Snake}\\ {Flower}\end{tabular}} & \multirow{2}{*}{} & \multirow{2}{*}{\begin{tabular}[c]{@{}l@{}}Flower\\ Fish\end{tabular}} & \multirow{2}{*}{\begin{tabular}[c]{@{}l@{}}Snake\\ Fish\end{tabular}} & \multirow{2}{*}{\begin{tabular}[c]{@{}l@{}}Flower\\ Carrot\end{tabular}} &  & \multirow{2}{*}{\begin{tabular}[c]{@{}l@{}}Flower\\ Carrot\end{tabular}} \\
& & & & & & & & \\
\bottomrule
\end{tabular}
\end{table*}

\textbf{Collaborative Story-making Activities:}  The eight children were arranged into four pairs. Each pair of students selected two to three preferred characters they wanted to enact in their stories. 
Children were given the freedom to choose with their partner from six basic and easy-to-understand emotions \cite{doi:10.1177/1077800414550462} such as "Happy," "Sad," or "Angry", and embody the selected emotion(s) in their stories.

To explore children's preferences for emotional responses in designing a child-oriented agent, the participants were invited to hold the paper cutouts of their played characters and role-play with each other in the collaborative story-making. They are also encouraged to combine other physical materials and make new artifacts (e.g., via collaborative drawing, as shown in \autoref{fig:teaser}). 
Guided by a construction model inspired by therapeutic storytelling \cite{perrow2008healing}, the model following a three-part framework of ‘metaphor’, ‘journey’, and ‘resolution’. Children, with researcher facilitation, constructed unique story plots using drawing, role-playing, and narratives. 
The deliberately open structure allowed for ample creative expression, encouraging children to proactively process emotions, demonstrate empathy, and establish positive relationships in the storytelling process \cite{jones2017promoting}. 
The right side of \autoref{fig:teaser} visually illustrates a typical social-emotional story created by the children during this process.

\textbf{Collaborative Story-Sharing Activities:} Similar to the last activity in Co-design Study 1, children in Co-design Study 2 were again arranged in a row to share and reflect on the collaborative stories they co-created. Each group actively engaged in acting out the stories using both verbal and non-verbal communication. The combination of these communication modes allowed children to embody the stories, providing a more immersive and expressive experience. This approach is inspired by the idea that acting out stories facilitates a deeper connection with the narrative \cite{rice2018story}. These emotional events were also designed as personal metaphors, enabling children to project their thoughts and feelings in a nuanced and imaginative manner \cite{perrow2008healing}.

\section{Findings}

\subsection{Desirable Characteristics and Features of the Agents}
In the designs created by children, some distinct patterns about their preferences of agent behaviors and appearance were observed. Familiar items or creatures were often chosen by the children as their characters like a flower, pig, carrot, or panda. They were also given whimsical (P: “I want a pink flower with pink feet and a big smile”) and imaginative features such as color-changing abilities for the snake and pig wearing ``orange glasses and shoes''. These differing choices reflect a desire for fun and playfulness, while also highlighting the highly diverse individual preferences of children. This indicates the importance of vibrant, imaginative design with abundant space for user-customization or user-authoring in capturing children's interest. 

These characters indicate a balanced approach to entertaining and supporting children emotionally: ``[My lollipop dropped…] Don’t worry, let’s buy you a new one!''. In addition, the characters take a nurturing and caring role, as a colorful snake that ``changes colors and tells stories at night to soothe it to sleep'', providing comfort and aiding in sleep. Color variability in the agent's designs and different activities mentioned in the stories (e.g., playing games together, going out for dinner, visiting the beach), again, suggest a tendency toward personalization where agents are tailored to the children's personalities and interests. Furthermore, many designs included real-life roles (e.g., friends, bullies, companions), contexts, and advice such as promoting healthy eating habits (``Eat more spinach, it’s good for your health, I also like to eat vegetables''), reflecting a preference for agents that are practical and helpful in children's life.

\subsection{Emotional events co-created by children}
After two co-design activities, researchers analyzed co-design video recordings, categorizing 14 social-emotional events generated by children (Detailed events can be found in the APPENDIX), with events classified into four types: Bullying, Recreation, Injuries, and Disputes.

Bullying incidents involve the theft or destruction of the child's belongings. Recreation features the agent character engaging in entertaining activities, like swinging at the park or building a snowman after dinner. Injury events include actions like twisting the foot, breaking the head, and falling. Disputes arise from disagreements, like conflicting weather preferences. Such social-emotional occasions created and enacted by children are likely to happen in real life and reflect children's responses to real-world experiences \cite{woolley2007development}. 

Emotional transformations can be observed in various stories created by children, reflecting an interesting pattern. Notably, we did not prompt these transformations and they emerged spontaneously in children's co-creation. The majority of shifts involve a transition from negative to positive emotions, such as moving from sadness to happiness (e.g., a broken lollipop replaced with a new one) or from anger to happiness (resolving conflicts and going out to play), which helps children learn emotional management and adjustment through storytelling \cite{pulimeno2020children,aznar2013spanish}. A few instances include emotions shifting from positive to negative, like happily going out shopping but ending up with a foot sprain and feeling sad. Additionally, a unique case that is in the loop of injury and play involves the agent experiencing multiple transformations within a single emotional event, adding complexity to the observed emotional dynamics.

\subsection{Emotional responses in the co-creative stories}
In the co-created stories, characters' social-emotional responses to each other can be categorized into two distinct types: verbal and non-verbal responses. These responses are indicative of various emotions enacted during the collaborative process.

\textbf{Verbal responses:}  In children's co-created stories, verbal responses express a spectrum of emotions. Instances of sadness are conveyed through descriptions of events that evoke such emotions, like damage or harm (e.g., ``The little flower whispered'' ``My candy stick got all squished up!'' or ``The little fish said: 
I got bumped and it hurt when I was playing in the park.''). 
Joyful expressions emerge through enthusiastic narrations of enjoyable situations (e.g., ``The little fish said: The warm sea baths feel super cozy! Yay!'') and discussions around play invitations (e.g., ``The little flower suggested: Hey Carrot, wanna go play at the park with me?, and the carrot excitedly said `Yay, yay, yay!'''). Angry emotions find expression in intense quarrelsome conversations and vivid depictions of events causing anger. Verbal responses from the partner character often extend to acts of care, comfort, and proposed solutions. For instance, when The little flower's lollipop was smashed, The little fish comforted her, promising to get her another one (``It's okay, I'll get a new one. Next time, keep your lollipop safe, okay?''). Urgent expressions surface in anxious moments, encouraging haste (``Hurry up! Why are you walking so, so slow?''). This nuanced exploration of verbal responses highlights the rich emotional tapestry woven into collaborative story-making among children.

\textbf{Non-verbal responses:} Similar to verbal responses, non-verbal expressions play a crucial role in conveying various emotions. We've categorized non-verbal responses based on each emotion, unveiling distinctive patterns. In moments of happiness, the agent's non-verbal expressions involve spontaneous behaviors (``The little flower is laughing happily,'' ``The little flower is jumping and walking, and the little snake is twisting and walking'') and interactive behaviors aimed at sharing joy with others (e.g., ``ticking the carrot, trying to make the carrot happy,'' or ``going out and having fun while holding the hand of the little fish''). In instances of sadness, the agent's non-verbal responses include making crying sounds and protective gestures, such as holding injured body parts (e.g., holding an injured knee, cradling a broken head, or wiping tears). In emotions like anger, various non-verbal responses manifest in spontaneous or interactive behaviors, including physical actions like mutual aggression during anger and fidgeting due to shyness. Spontaneous behaviors, such as opening hands in response to stimulation, falling asleep, nodding, and body-stretching, occur when comforted. Interactive behaviors include gentle patting, head-touching, embracing, and handshakes to soothe others. This comprehensive classification captures the diverse range of non-verbal expressions associated with different emotions exhibited by the agent during collaborative story-making.

\section{Discussing Design Implications}

\textbf{Implication 1: Fostering emotional sharing between children and agents through prompting inquiries within the dialogue.} In this context, ``prompting inquiry'' refers to encouraging a child to ask about the emotional status of an agent, or vice versa. An illustrative example from the co-created stories is when one character cries due to injury or bullying, prompting the other to inquire about the situation. The agent's proactive questioning becomes a catalyst for empathetic conversation. Conversely, when children perceive emotional expressions from the agent, it provides them with an opportunity to learn from fictional social-emotional scenarios. Integrating narrative and dialogue into curricular planning, as advocated by Piipponen et al. \cite{piipponen2019children}, enhances children's self-awareness and understanding of others through the sharing and listening of stories. This underscores the significance of this implication.

\textbf{Implication 2: Enabling the agent to narrate stories about its negative emotions, and collaborate with children to explore appropriate solutions.} For example, in a co-created scenario where the little flower, portraying the agent, quietly cries over a broken lollipop, the little fish approaches to offer comfort. the little fish reassures the little flower, suggesting the purchase of a new lollipop and advising on safeguarding it. In this process, the agent shares an emotional story, expressing a desire for children to contribute solutions. Through dialogue, children extend care and propose measures to address negative emotions, actively assisting the agent in coping with emotional challenges. This reciprocal interaction, akin to emotional role-playing games with robots discussed by Ji et al. \cite{ji2015design}, allows children to practice essential social skills for emotional learning. Moreover, children supporting agents experiencing negative emotions, in comparison with vice versa, could potentially safeguard children in SEL, against potential risks of children experiencing negative emotions directly.

\textbf{Implication 3: Leveraging various nonverbal elements, including vocal, auditory, or embodied actions to enrich the agent's emotional responses.} In our exploration of emotional responses in co-creative stories, we observed distinctive vocalizations like ``woo woo woo'' for sadness, cheering for happiness, and dubbing events within the story (e.g., mimicking the whirring sound of throwing snowballs or the beeping of an ambulance). Additionally, children often describe actions, like a flower opening its hands when joyfully swinging, to convey emotions, serving as valuable design references for incorporating non-verbal behaviors into the agent's interactions. Concrete emotional expression sound effects and movements, mirroring the diversity observed in children's nonverbal communication, have the potential to captivate children's attention and interest, fostering stronger interaction between the agent and children \cite{risnawati2021yufid}. This approach enriches multi-modal channels for emotional communication, potentially enhancing overall immersion and embodiment of SEL.

\textbf{Implication 4: Designing proper proactiveness of the agent.} In the co-created stories, the agent's active initiation takes various forms, such as inviting children to play in the park or sharing drinks. This proactive stance in establishing interactions contributes to the cultivation of friendship and a sense of belonging among children, increasing their willingness to engage with the agent. The deliberate effort to initiate varied interactions not only enriches the conversation experience but also reinforces the emotional bond between the agent and children, creating a more engaging and meaningful interaction.

\textbf{Implication 5: Discouraging children's undesirable behavior by thoughtfully designing emotional responses.} While the agent may not have the capability to detect inappropriate language or behavior outside of the dialogue context, it can leverage the emotional bond established during conversations. In response to inappropriate behavior or language, the agent can express sadness, utilizing the emotional connection to prompt the child to reflect internally. This approach encourages children to express their genuine emotions, facilitating introspection and self-correction. By incorporating mechanisms to discourage negative behavior within the dialogue, the agent contributes to creating a more respectful and positive interactive environment.

\textbf{Implication 6: Encouraging richer tangible interactions for children to communicate emotions to the agent.} In narratives crafted by children involving sad emotions, instances often arise where the agent, experiencing an injury, begins to cry, and the partner gently touches the affected part as a move to comfort. For example, if the little fish hugs his injured knee after a fictional beating or cradles a broken head, children can respond with more intricate tangible interactions like gentle pats, hugs, handshakes, or head touches to console the agent. Implementing this approach necessitates the development of an embodied agent capable of discerning the subtleties in children's movements, enhancing the depth and authenticity of emotional exchanges within the storytelling experience.

\subsection{Limitation and future work}
While our work provided interesting preliminary insights, as an early exploration, it has several limitations, including the limited sample size, lack of broad age distribution of child participants, lack of systematic, quantifiable analysis, and limited number of co-creation sessions. These limitations call for a larger-scale, longer-term, and mixed-method approach in our upcoming study. Additionally, the extended waiting time for the AIGC tool as experienced by children can be another aspect that we could improve, by adopting a faster tool in the future to enhance children's experience and engagement in co-creating. Our findings revealed children's sharing and enacting social-emotional stories, including sensitive topics like bullying, suggesting the potential for agents to serve as platforms for emotional caring beyond learning. With the intriguing pattern we observed, subsequent studies can delve deeper into the types of stories children prefer to share and extract preferred design qualities for emotional and developmental support and it will be valuable to explore how certain characteristics of AI, such as over-reliance, hallucinate results, and bias, affect the co-design process with children.

\begin{acks}
We appreciate all the collaborating teachers for their generous support from Sustech kindergarten. We thank all the participating children for their time and efforts in the co-design activities, as well as to their parents who supported them throughout the process. Our thanks also go to the reviewers for their invaluable comments that helped us significantly improve this paper. This work is supported by the National Social Science Fund of China (NSSFC) Art Grant: No. 22CG184.
\end{acks}


\bibliographystyle{ACM-Reference-Format}
\bibliography{sample-base}

\appendix

\section{appendix}
\begin{figure*}[ht]
  \centering
  \includegraphics[width=15.6cm]{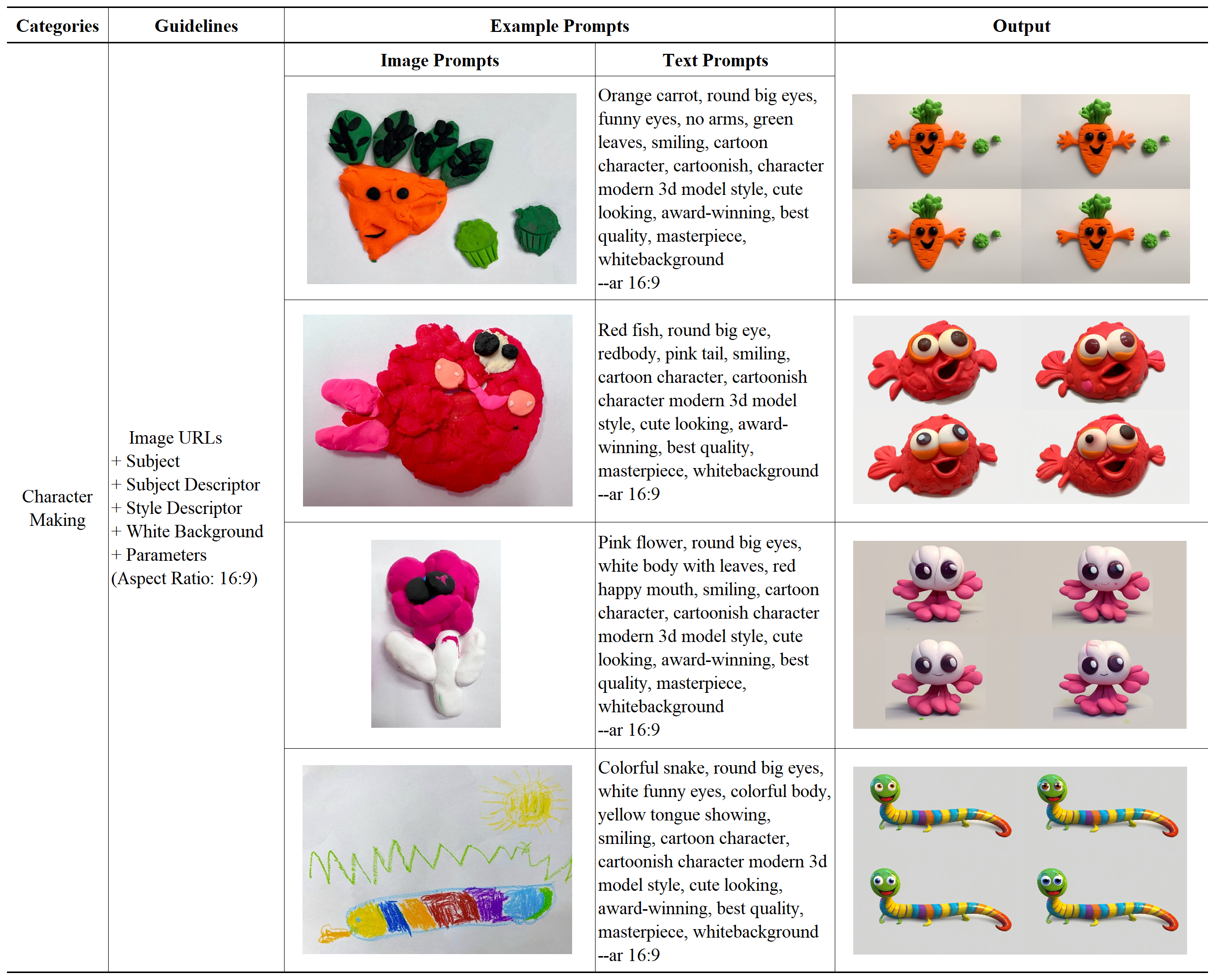}
  \caption{Guidelines and Examples on Midjourney Prompts.}
  \label{appendix}
\end{figure*}

\begin{table*}[ht]
\renewcommand{\arraystretch}{1.1} 
\centering
\caption{14 social-emotional events generated by children in Co-design Session 2}
\begin{tabular}{@{}>{\centering\arraybackslash}p{1cm}p{16cm}@{}} 
\toprule[\heavyrulewidth] 
\textbf{Number} & \textbf{14 social-emotional events generated by children} \\
\midrule
1               & "Candy Crushed: Receiving a Replacement Lollipop" \\
2               & "From Bullying to Care: Receiving Concern and Comfort" \\
3               & "Shopping Mishap: Twisted Ankle Rescued by a Caring Doctor" \\
4               & "Bumped Head: Bleeding and Healing at the Undersea Hospital with a Companion" \\
5               & "Hot Springs Delight: Joyful Adventures in the Ocean" \\
6               & "Clashing Views: Disagreements Leading to Arguments" \\
7               & "Playdate Adventure: Overcoming Shyness and Swinging on the Seesaw" \\
8               & "Mall Mishap: Injured, Hospitalized, and Visited by Kind Souls" \\
9               & "Post-Hospital Fun: Playing at the Mall Until Nightfall and Heading Home to Sleep" \\
10              & "Little Flower's Park Adventure: Inviting Radish to Swing, Enjoy Drinks, Bask in the Sun, Admire Dandelions, and Play with a Wild Rabbit" \\
11              & "The Squabble of Little Kid and Carrot: Rain or Shine Argument, Sun and Rain Take Sides, Carrot Concedes, Comforted by Little Kid" \\
12              & "Petty Dispute of Two Little Flowers: Reconciliation and Playtime Together" \\
13              & "Bumped Head: Comfort, Concern, Doctor's Visit, and a Proposal for Seesaw Fun" \\
14              & "Snowman Tumble After Dinner: Hospital Visit and Returning to Build Snowman" \\
\bottomrule[\heavyrulewidth] 
\end{tabular}
\end{table*}

\end{document}